\documentclass{aa} 
\makeatletter
\renewcommand*\aa@pageof{, page \thepage{} of \pageref*{LastPage}}
\makeatother

\newcommand{\kms}{km s$^{-1}$}
\usepackage{graphicx}
\usepackage{txfonts}
\usepackage[dvipsnames]{xcolor}

\usepackage{hyperref}
\colorlet{mylinkcolor}{BlueViolet}
\colorlet{mycitecolor}{BlueViolet}
\colorlet{myurlcolor}{BlueViolet}
\newcommand\myshade{85}
\newcommand\mjup{$M_{\rm Jup}$}
\newcommand\rjup{$R_{\rm Jup}$}

\hypersetup{
  linkcolor  = mylinkcolor!\myshade!black,
  citecolor  = mycitecolor!\myshade!black,
  urlcolor   = myurlcolor!\myshade!black,
  colorlinks = true,
}

\begin{document} 

   \title{ALMA and NACO observations towards the young exoring transit system J1407 (V1400 Cen)}

   \author{ M.A. Kenworthy
          \inst{1}
          \and
          P.D. Klaassen
          \inst{2}
          \and
          M. Min
          \inst{3}
          \and
          N. van der Marel
          \inst{4,5}
          \and
          A.J. Bohn
          \inst{1}
          \and
          M. Kama
          \inst{6}
          \and
          A. Triaud
          \inst{7}
          \and
          A. Hales
          \inst{8,9}
            \and
          J. Monkiewicz
          \inst{10}
          \and
          E. Scott
          \inst{11}
          \and
          E.E. Mamajek
          \inst{12,13}
          }

   \institute{Leiden Observatory, Leiden University, P.O. Box 9513, 2300 RA Leiden, The Netherlands \\
              \email{kenworthy@strw.leidenuniv.nl}
              \and
              UK Astronomy Technology Centre, Royal Observatory Edinburgh, Blackford Hill, Edinburgh EH9 3HJ, UK
              \and
              SRON Netherlands Institute for Space Research, Sorbonnelaan 2, 3584 CA Utrecht, The Netherlands
              \and
              Herzberg Astronomy \& Astrophysics Programs, National Research Council of Canada Herzberg, 5071 West Saanich Rd, Victoria, BC, V9E 2E7, Canada
              \and
              Department of Physics \& Astronomy, University of Victoria, Victoria, BC, V8P 1A1, Canada
              \and
              Institute of Astronomy, University of Cambridge, Madingley Road, Cambridge, CB3 OHA, UK
              \and
              School of Physics \& Astronomy, University of Birmingham, Edgbaston, Birmingham B15 2TT, UK
              \and
              National Radio Astronomy Observatory, 520 Edgemont Road, Charlottesville, VA 22903-2475.
              \and
                Atacama Large Millimeter/Submillimeter Array, Joint ALMA Observatory, Alonso de Cordova 3107, Vitacura 763-0355, Santiago
              \and
              School of Earth and Space Exploration, Arizona State University, Tempe, AZ 85287, USA
              \and
              Chinese Academy of Sciences, 52 Sanlihe Road, Xicheng District, 100864 Beijing, China
              \and
            Jet Propulsion Laboratory, California Institute of Technology, M/S 321-100, 4800 Oak Grove Drive, Pasadena, CA 91109, USA
            \and 
            Department of Physics \& Astronomy, University of Rochester, Rochester, NY 14627, USA}

   \date{Accepted 2019 November 16}

\titlerunning{J1407 with ALMA and NACO}

 
  \abstract
  {} 
  {Our aim was to directly detect the thermal emission of the putative exoring system responsible for the complex deep transits observed in the light curve for the young Sco-Cen star 1SWASP J140747.93-394542.6 (V1400 Cen, hereafter J1407), confirming it as the occulter seen in May 2007, and to determine its orbital parameters with respect to the star.}
  {We used the Atacama Large Millimeter/submillimeter Array (ALMA) to observe the field centred on J1407 in the 340 GHz (Band 7) continuum in order to determine the flux and astrometric location of the ring system relative to the star.
  We used the VLT/NACO camera to observe the J1407 system in March 2019 and to search for the central planetary mass object at thermal infrared wavelengths.}
    {We detect no point source at the expected location of J1407, and derive an upper limit $3\sigma$ level of $57.6~\mu\rm{Jy}$.
      There is a point source detected at an angular separation consistent with the expected location for a free-floating ring system that occulted J1407 in May 2007, with a flux of $89~\mu\rm{Jy}$ consistent with optically thin dust surrounding a massive substellar companion.
      At 3.8 microns with the NACO camera, we detect the star J1407 but no other additional point sources within 1.3 arcseconds of the star, with a lower bound on the sensitivity of $6M_{Jup}$ at the location of the ALMA source, and down to $4M_{Jup}$ in the sky background limit.
      }
   {The ALMA upper limit at the location of J1407 implies that a hypothesised bound ring system is composed of dust smaller than $1\rm{~mm}$ in size, implying a young ring structure.
   The detected ALMA source has multiple interpretations, including: (i) it is an unbound substellar object surrounded by warm dust in Sco-Cen with an upper mass limit of $6M_{Jup}$, or (ii) it is a background galaxy.
   %
}

   \maketitle
%

\section{Introduction}

Planet formation theory predicts that gas giant planets form in circumstellar disks (CSDs) which disperse on a timescale of $2-3$ Myr \citep{Williams11}. 
Once a planetary embryo forms within the CSD, a circumplanetary disk (CPD) builds inside the gravitational domain of the embryo, and gas and dust are transported through this disk and accrete onto the planet \citep[see reviews by e.g.][]{Armitage11,Kley12}.
Once the CSD disperses, the remaining gaseous CPD accretes onto the planet in a short time \citep{Szulagyi14,Perez15} or possibly on longer timescales depending on the planet formation mechanism \citep{Szulagyi17} and may form moons within the Hill radius of the planet \citep{Canup02}.
The duration of this intermediate stage is not strongly constrained by observations \citep[upper limits of millimetre-sized particles to CPDs around known exoplanets are reported in][]{Perez19}, but the large solid angle subtended by the Hill sphere filling disk and rings imply that transits of these systems may be common around young stars.

1SWASP J140747.93-394542.6 (hereafter J1407) is a pre-main-sequence, $\sim 16$ Myr old, $0.9M_\odot$,$V=12.3{\rm~mag}$ K5 star with parallax $7.1835\pm 0.0447{\rm~mas}$ \citep{Gaia18}, consistent with  $d=138.7\pm 0.6{\rm~pc}$ \citep{Bailer-Jones18} and membership of the Sco-Cen OB association \citep{Mamajek12}.
\citet{Mamajek12} reported  light curves for J1407 from the Super Wide Angle Search for Planets \citep[SuperWASP;][]{Butters10} and the All Sky Automated Survey \citep[ASAS;][]{Pojmanski02}.
These light curves show that the star underwent a complex series of deep transits lasting 56 days around 2007 May and includes a maximum dimming of >95\%.
The gradient of the light curve during the eclipse in 2007 can be used to derive a transverse velocity of about 35 \kms{}; combined with the 56 day duration, this leads to a diameter of the order of 1.2\,au, marginally consistent with a Hill sphere filling structure around a substellar companion to J1407.
An inversion of the light curve suggests several dozen rings with at least one cleared gap suggesting the presence of a forming exomoon \citep{Mamajek12,vanWerkhoven14,Kenworthy15b}.
Subsequent papers have detailed the search for successive eclipses in photometric data \citep{Kenworthy15b,Mentel18} through radial velocity and direct imaging with full aperture and non-redundant mask imaging \citep{Kenworthy15} that have placed successively more  stringent limits on the mass and orbital period of the putative companion.
Circular orbits are strongly ruled out with these studies and the diameter of the ring system model, along with the detailed small-scale structure deduced from nightly flux variations, makes it very challenging to explain the apparent stability and coherence of the rings \citep{Zanazzi17,Rieder16}.
Another challenge is whether the ring gaps are caused by mean-motion resonances (MMRs) or by direct clearing, and the elliptical orbit at periastron disfavours the MMR hypothesis \citep{Sutton19}.
With the central massive object undetected, we turn to detecting the thermal radiation from the rings themselves.
The rings are estimated to be 1.2\,au in diameter based on the model described in \citet{Kenworthy15b}.

In this paper we present data from the Atacama Large Millimeter/submillimeter Array (ALMA) taken in 2017 in a search for thermal emission from the rings of the J1407 occulting object.
We report a detection of a source consistent with millimetre-sized material close to the location of the star with ALMA, but we detect no point source at thermal infrared wavelengths at the location of the ALMA source.
The non-detection in the thermal infrared is consistent with an upper mass limit of $6M_{Jup}$ at the age of the Sco-Cen association.
The separation of the detected source from the location of J1407 is consistent with the location of a free-floating object with rings that has moved at a constant velocity (derived from the original eclipse with a value of 35 \kms{}) since the eclipse seen in 2007.

In Sections \ref{alma} and \ref{naco} we describe the observations with ALMA and NACO and the subsequent data reduction.
The interpretation is dependent on the astrometry, and this is discussed in detail in Section~\ref{astrom}.
In the last sections we discuss the hypotheses for this detection and future observations that will constrain the nature of this source.

\section{Observations with ALMA\label{alma}} 

The star J1407 was observed by ALMA on UT 20, 22, and 24 July 2017 for a total of 2 hours on source and a total time of  5.11 hours.
Bandpass, phase, and flux calibrations were all carried out using J1427-4206, which had a flux between 1.46 and 1.5 Jy in the observed spectral windows (at 336.495, 338.432, 348.495, and 350.495 GHz).
The observations were made in time domain mode (TDM) with 128 channels per spectral window.
The mean atmospheric precipitable water vapour (pwv) content was between 0.5 and 0.86 mm across the observations, as summarised in Table \ref{tab:Obs_params}.
Similarly, the minimum and maximum baseline lengths for each observation is presented in the same table. 

\begin{table*}
\caption{Calibrators and observing parameters for ALMA project 2016.1.00470.S.}
\begin{tabular}{lllllllll}
\hline \hline
&Time on & \multicolumn{3}{c}{Calibrators} & & \multicolumn{2}{c}{Baseline Lengths} & Mean \\ \cline{3-5} \cline{7-8}
Execution  & Science Source & Phase & Bandpass & Flux & Observing Date & Max & Min & PWV\\
 & (min) &&&&&(m) &(m)\\
 \hline
X54fd & 40 & J1427-4206 & J1427-4206 & J1427-4206 & 2017-07-24 & 3638 & 30.5 &0.51\\
X238e & 40 & J1427-4206 & J1427-4206 & J1427-4206 & 2017-07-20 & 3697 & 16.7 &0.86\\
Xd6d & 40 & J1427-4206 & J1427-4206 & J1427-4206 & 2017-07-22 & 3697 & 16.7 &0.50\\
\hline
\end{tabular}
\label{tab:Obs_params}
\end{table*}

All ALMA data in this paper were reduced using {\sc casa} 4.7.2 \citep{McMUllin07}.
The data were originally pipeline reduced, with a secondary phase calibrator (J1405-3907) used in a pre-calibration gain table to aid the gain calibration of the primary phase calibrator.
However, this secondary calibrator was later found to be a double source.
This could increase the uncertainty of the phase calibration, and thus the astrometric uncertainties.
To control for this uncertainty, the double source secondary calibrator was flagged out of the dataset and explicitly removed from the calibration process (in the first \texttt{hifa\_flagdata()} pipeline step), and the pipeline was re-run without it.
The calibration and final products were unchanged, but any astrometric uncertainty introduced by having an extended phase calibrator was removed.
Self-calibration of the final dataset was not possible as the target was too faint.
The final calibrated data was imaged using natural weighting, which resulted in a synthesised beam of 0.12$\times$0.07$''$ at a position angle of -77.5$^\circ$.

\begin{figure}[tbh]
\begin{center}
\includegraphics[width=\columnwidth]{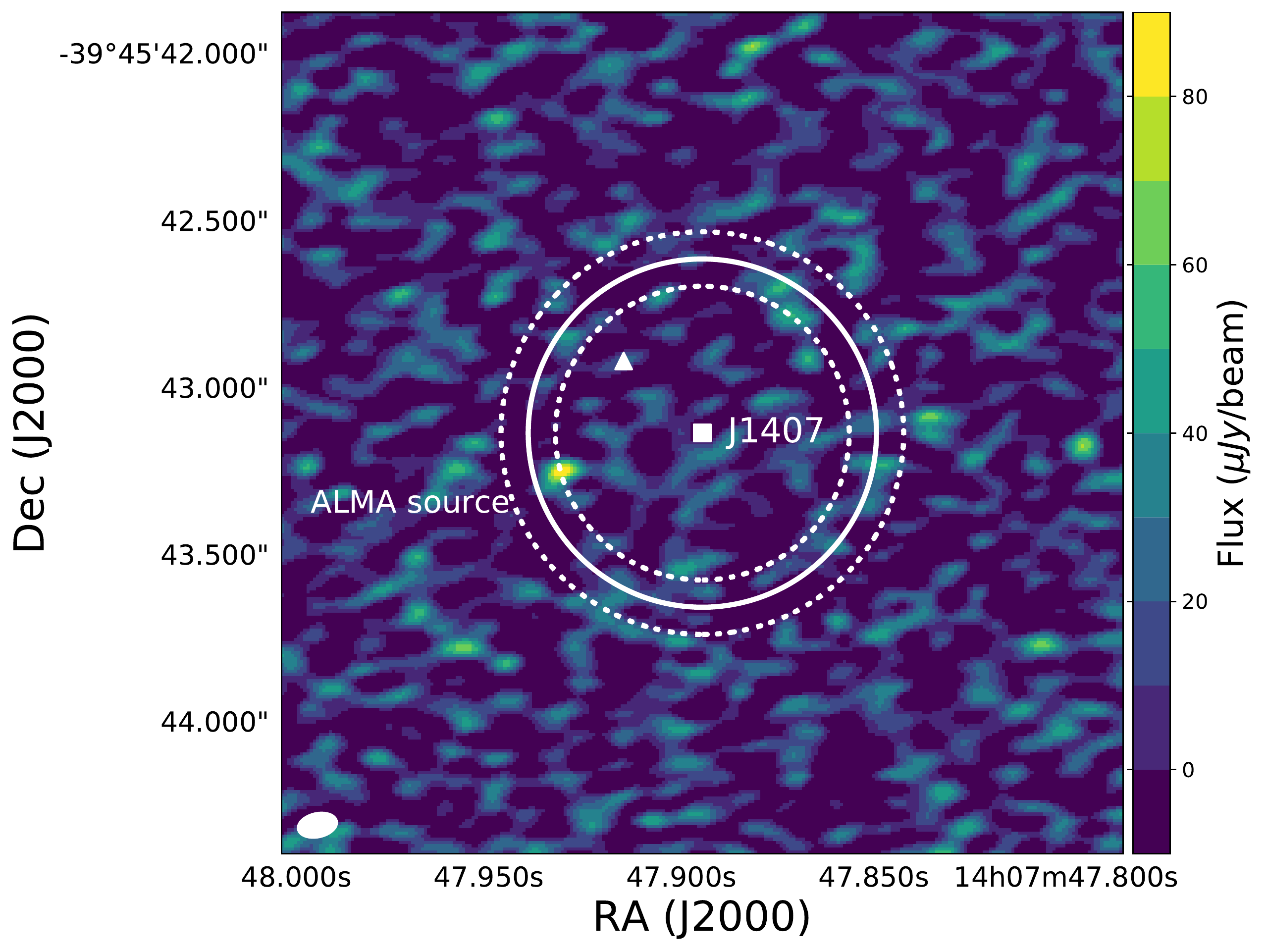}
\caption{\label{fig:alma}ALMA image of the field towards J1407. The location of the star J1407 in 2017 is shown as a white square, and the white triangle gives the location at the time of the eclipse in 2007. The white ring indicates the expected location for an object travelling at 35 \kms{} between the two epochs (assuming that the object is co-distant with J1407);  the dotted lines represent the 1$\sigma$ uncertainties. The detected ALMA source is at a position angle of approximately 95 degrees. The ALMA beam is shown in the lower left corner.}
\end{center}
\end{figure}

A point source was detected in the ALMA field near the expected location of J1407 (see Figure~\ref{fig:alma}).
As the point source was near the phase centre of the observations, we did not do any primary beam correction.
Using the {\sc casa} task \texttt{imfit}, we find that the source has a position of $\alpha$ = 14:07:47.9311($\pm$0.00062), $\delta$ = -39:45:43.26068($\pm$0\arcsec.00388) (ICRS), and a size of (95.1\,$\pm$\,18.1) $\times$\, (58.4\,$\pm$\,6.1)\,mas at a position angle of 113$^{\circ}$.4\,$\pm$\,8$^{\circ}$.4, which means it is unresolved in our synthesised beam.
Using the {\sc casa} task \texttt{imstat}, we derive a peak flux of 89 $\mu$Jy/beam and an rms noise of 19 $\mu$Jy/beam, resulting in a 4.7 $\sigma$ detection.
We take the J1407 stellar parameters to be $T_* = 4500 K$, $R_* = 1.039\pm 0.075 R_\odot$, and $L_* = 0.40\pm0.02 L_\odot$ at a distance $d = 139$ pc and a frequency of 345 GHz.
The stellar photosphere flux $F_*$ from J1407 is therefore  

    $$F_* = B_\nu(T_*) \pi R_*^2/d^2 = 1.5\,{\rm \mu Jy}$$
    
This is much fainter than the detection limit of our observation, and so the star is not expected to be visible in our image.

\section{Observations with NACO\label{naco}} 

The star J1407 was observed with VLT/NACO \citep{Lenzen03,Rousset03} on 1 March 2019 UT at a wavelength of $3.8\mu m$.
The observing conditions were excellent with an average seeing of $0\farcs46\pm0\farcs08$ and a coherence time of $5.66\pm1.12$ milliseconds.
The data were obtained in cube burst mode, with telescope pupil tracking \citep{Kasper09} enabled to minimise the effect of residual speckles from the telescope optics for optimal Angular Differential Imaging \citep[ADI;][]{Marois06}.

Excluding telescope and instrument overheads the on-source integration was 5\,220 seconds with a field rotation of 71\fdg8.
The star J1407 was unsaturated in the images and acted as a PSF reference for subsequent data reduction and fake companion injection.
Processing the data with the latest release of the \texttt{PynPoint} package \citep{Stolker19} and applying Principal Component Analysis \citep[PCA; ][]{Amara12} results in a sensitivity map for point sources.
Close to the star, the sensitivity is dominated by noise from the stellar speckle halo, falling to the sky background level noise limit at larger angular separations.
The processing is dependent on the number of PCA components used in the data reduction pipeline.
A larger number of PCA components removes more speckles at small angular separations, but also removes flux from any point sources in the image.
To correct for this systematic loss of companion flux, fake point sources are injected at several radii from the star into the data, and then the recovered flux is measured after the pipeline data reduction.
The 5 $\sigma$ contrast was calculated by injecting fake companions (created from the stellar PSF) at various angular separations (0.14 to 1.20 arcsec in 0.054 arcsecond steps) and position angles (at 60 degree increments starting from 0 degrees).
The small number of statistically independent patches close to the central star means that a correction is required to the sensitivity as described in \citet{Mawet14}, and we apply this correction to the contrast curve.

\begin{figure}[tbh]
\begin{center}
  \includegraphics[width=\columnwidth]{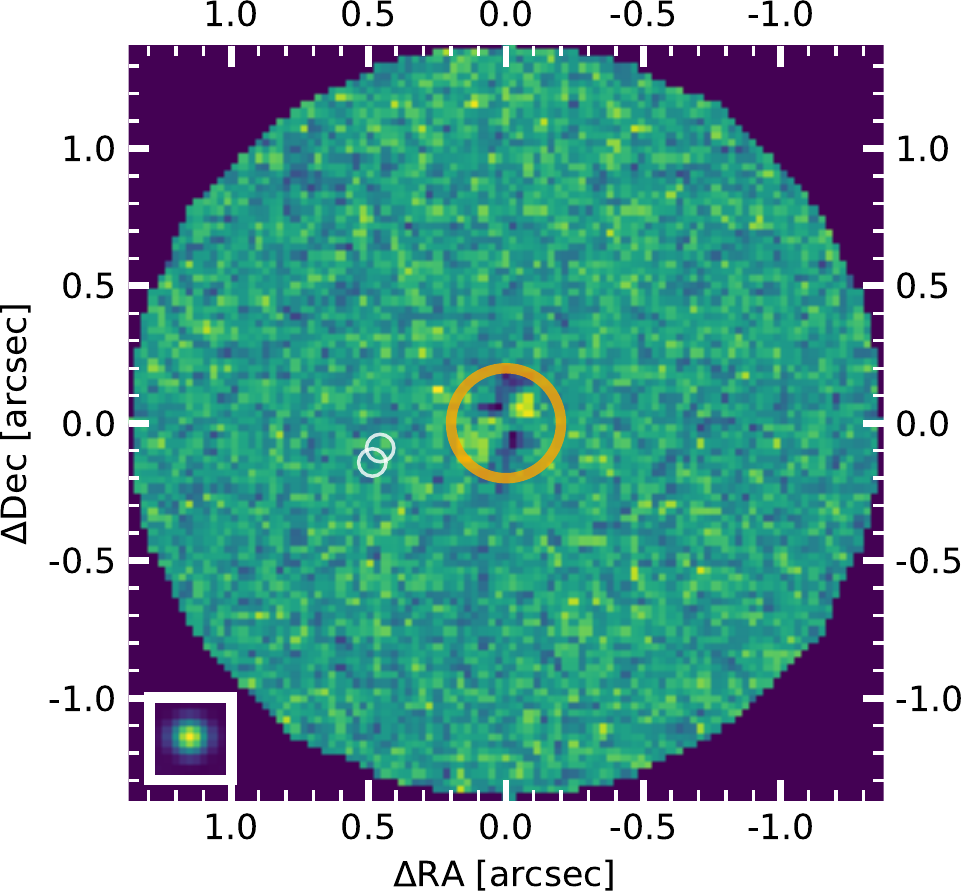}
\caption{\label{fig:naco}Image at L' band towards J1407. The star is in the middle of the field of view. The two white circles indicate the location of the ALMA source if it is a distant unrelated source and if it has proper motion consistent with an eclipse in 2007. Systematic errors and loss of point source sensitivity occurs within 0.2 arcseconds of J1407 (orange circle). No significant point sources are detected around the star.}

\end{center}
\end{figure}

We determined that 5 PCA components is a reasonable compromise between speckle removal and sky background sensitivity, and the final image is shown in Figure~\ref{fig:naco}, which shows that the sensitivity is dominated by the noise from the sky background in all but the central 0.3 arcseconds.

\begin{figure}[tbh]
\begin{center}
  \includegraphics[width=\columnwidth]{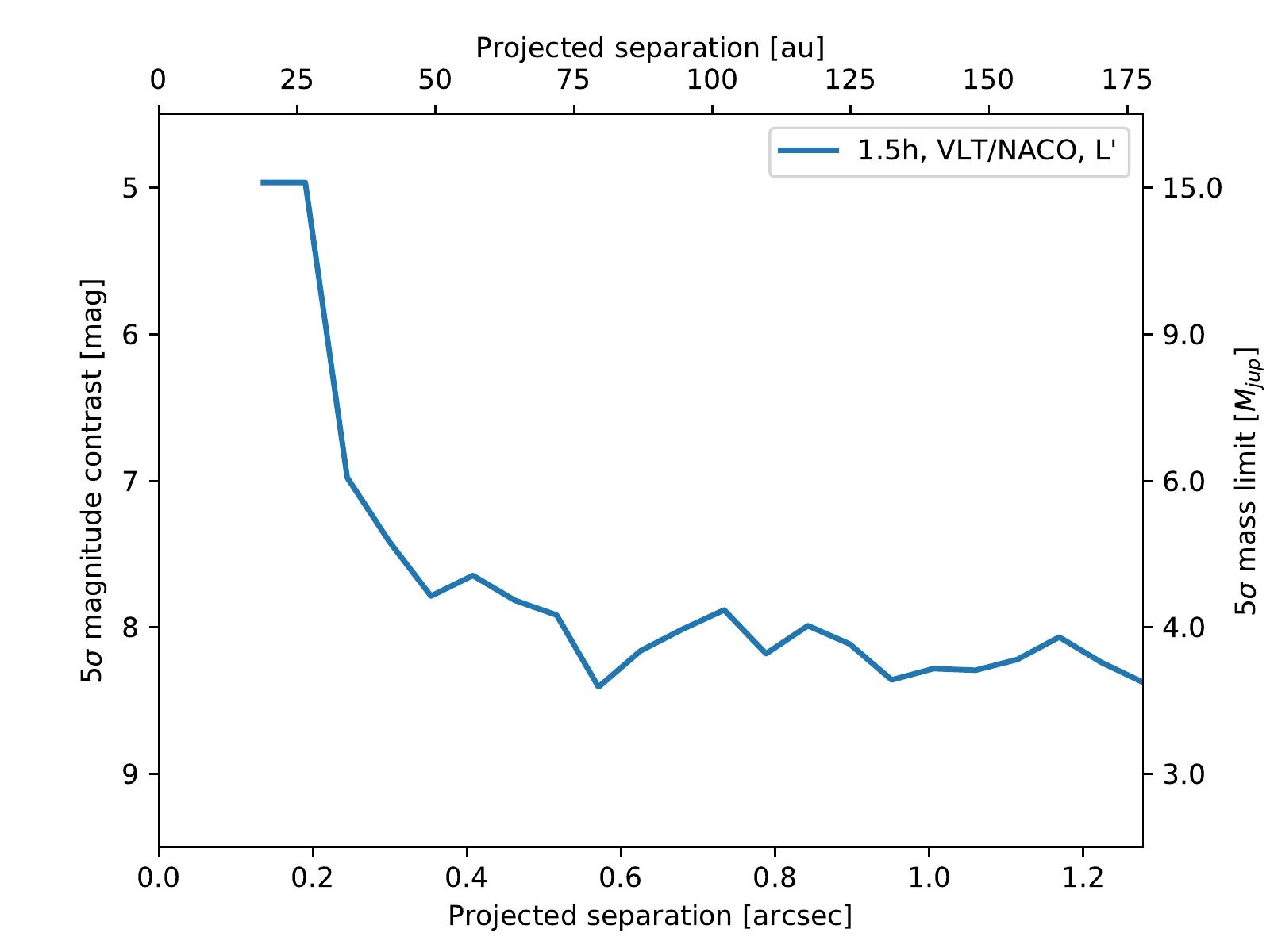}
\caption{\label{fig:contrastc}Contrast curve for NACO L' observations towards J1407. The curve shows the $5\sigma$ point source limit and uses the AMES-Cond model to convert from apparent magnitude to planetary mass.}

\end{center}
\end{figure}

To convert from sensitivity in magnitudes to companion mass, we use the  AMES-Cond \citep{Baraffe03,Allard01} models for conversion of contrast to Jupiter masses and assume an age of 16 Myr corresponding to the age of Sco-Cen, with the resultant contrast curve shown in Figure~\ref{fig:contrastc}.
No point sources are detected within the field of view, down to $6M_{Jup}$ at 30au (0.25 arcsec) and dropping to less than $4M_{Jup}$ beyond 100au (0.70 arcsec).

\section{Astrometric analysis}\label{astrom} 

There is no source detected with ALMA and NACO within 100 milliarcseconds of the star J1407, which would be consistent with a companion in a bound orbit, as determined by the 2007 eclipse and subsequent non-detections of a repeating eclipse event \citep{Mentel18}.
The $3\sigma$ upper limit on this ALMA non-detection is $57.6{\rm~\mu Jy}$.
The detected source is too far away from the star to be gravitationally bound to the star (the escape velocity from the star at this separation is a small fraction of the derived transverse velocity), but its position is marginally consistent with a free-floating object that transited J1407 in 2007 and is moving with a transverse velocity of 35 km\,s$^{-1}$ relative to J1407 (assuming source is co-distant with J1407) derived from the 2007 observations \citep{Kenworthy15b}, as indicated by the white circle in Figure~\ref{fig:alma}.
Upper and lower uncertainties on the transverse velocity location are shown by the dashed circles.
The Sco-Cen association and its subgroups are unbound \citet{Wright18} with a 3D velocity dispersion in UCL region of 2.45 \kms{}, leading to an implied 1D velocity dispersion of 1.4 \kms{} at the location of J1407, meaning that this object is unbound from the association and will leave it on a timescale of several million years, presumably having been ejected from one of the stellar systems in the association.

The measured brightness of the point source is consistent with our models for a ring system with a ring particle size greater than 100 microns (see next section), but the distance between the ALMA source and the location of J1407 in 2007 is $438\pm 8$ mas, which is equal to a separation of 61~au at the distance of J1407.
We have spent the months since the observations confirming with both the ALMA Allegro node in Leiden and independently with our own reduction pipelines that the astrometry is consistent, and so we adopt this astrometric solution   for the rest of the paper.

\section{Modelling of a ring system at millimetre wavelengths} 

The optical eclipse of 2007 was seen at broad-band optical wavelengths, with no colour information obtained during the eclipse.
Analysis of the optical light curve show sharp transitions in optical depth, implying the presence of discontinuities in particle density with ring radius, very similar to the rings of Saturn.
This indicates that the rings are geometrically thin \citep[with a height-to-diameter ratio $h_{ring}/d_{ring}\approx 0.001$;][]{Mamajek12}
We  note that this estimate is only valid where large gradients in the light curve are seen;  during the time of smallest projected separation of J1407 behind the disk system, the derived height-to-diameter ratio can be much higher as the path of the star is then tangential to any azimuthally symmetrical structure in the disk (see Figure 3 in \citealt{Kenworthy15b}).
Both of these findings point to a system in which the relative velocities between the particles may be small, leading to dust collisions that are not destructive but build up to larger sizes.
This is also seen in Saturn's rings where the dust particles are millimetre to centimetre sized \citep[e.g.][]{Tiscareno13}.
To model the continuum emission from the ring system, we adopt   the model from \citet{Baraffe08}, which is closest to the mass and age constraints and where the ring-hosting object has a mass $5\,$\mjup, age $16.67\,$Myr, radius $1.337\,$\rjup, and
luminosity $\log_{10}(L_{\rm planet}/{\rm L_{\odot}})=-4.69$.


The flux from the ring system is dependent on the size distribution of the dust particles.
We model the dust in the ring system using a 3D version of the radiative transfer code MCMax \citep{Min09} for several different input grain size distributions.
Models for the luminosity of the central substellar object are taken from \citet{Baraffe08} for an age of 16~Myr.

{\bf Surface density distribution:} We assume an outer radius of 0.6 au for the ring system (equivalent to the Hill radius derived above) and the number density of the dust particles is adjusted to match the optical depth measured at optical wavelengths for each of the rings from the model in \citet{Kenworthy15b}.
The surface density is calculated to match the optical depth in the rings as measured: $\Sigma = \tau_V/\kappa_V$ with $\Sigma$ the surface density, $\tau_V$ the optical depth in the visible, and $\kappa_V$ the mass absorption cross section in the visible.
The opacities are computed using irregularly shaped grains \citep[applying the method by][]{Min05} composed of a mixture of silicate and ice (40\% Mg rich pyroxene, 60\% water ice).
This is consistent with the amount of water and silicate that  would form from a mixture of elements with solar composition, using up all the Si, Mg, and O.
We note that we do not model any dust at radii smaller than 0.19 au, which was the smallest projected separation derived in \citet{Kenworthy15b}.
Our model represents a lower bound on the amount of dust available for heating in this system.
The parameter space for ring models is explored by varying the separation of the ring system from the star (from 5 au to 50 au), varying the luminosity of the substellar object at the centre of the rings, and using several different configurations for the dust in the rings. 

\begin{figure}[tbh]
\begin{center}
\includegraphics[width=\columnwidth]{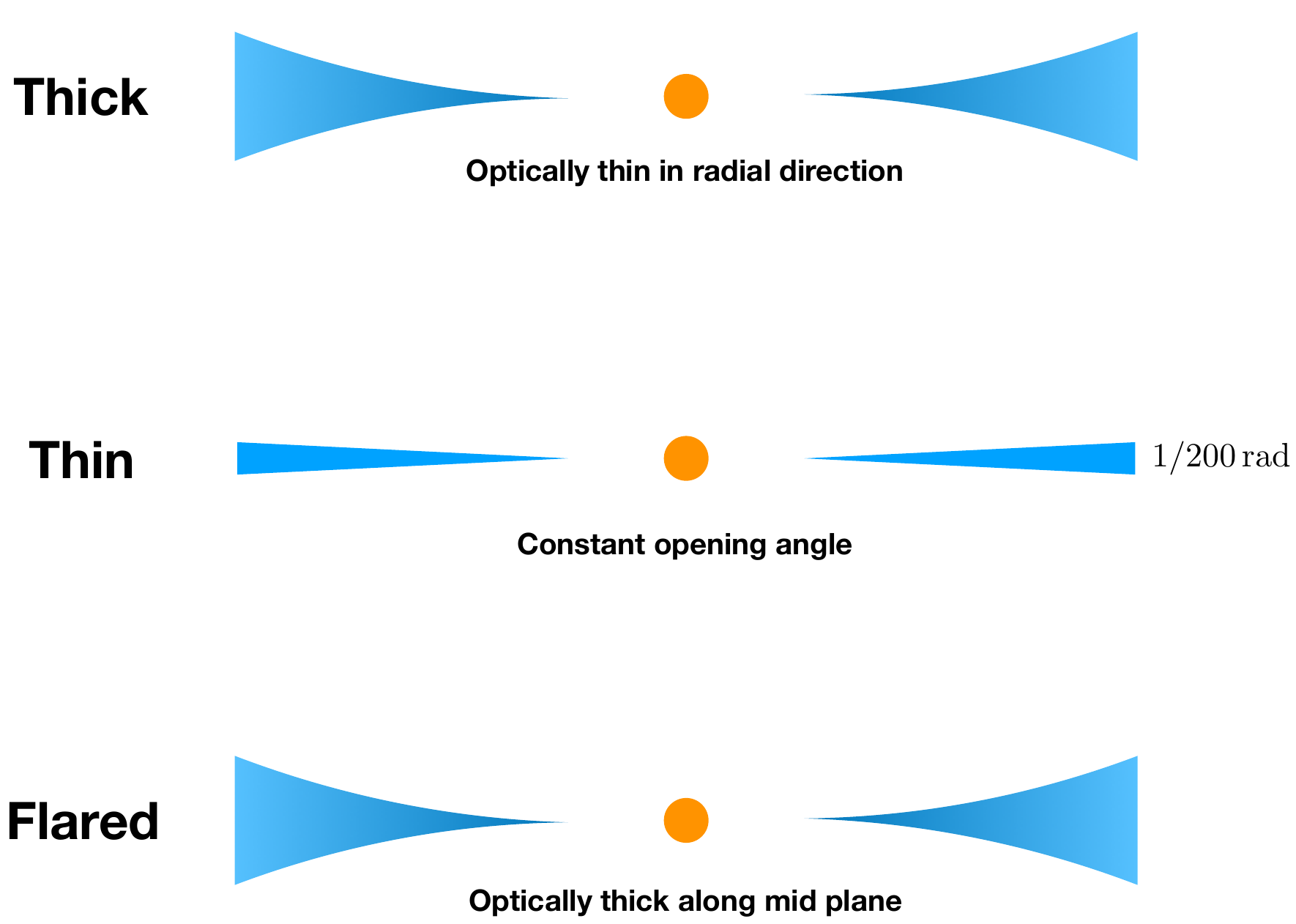}
\caption{\label{fig:diskmodels}Diagrams of the three models for the dust (blue) around a central low mass substellar companion (orange). See text for detailed descriptions of the models.}

\end{center}
\end{figure}

{\bf Vertical distributions:} We chose three configurations of the distribution of dust around the central object and these are shown in Figure~\ref{fig:diskmodels}.
The `Thin' model has a disk whose scale height scales linearly with $R$ as $H=0.005 R\rm{~au}$, so that along the midplane of the disk the dust is optically thick and self-shields flux from the central object.
This is to test the hypothesis that the disk is dynamically cold and highly coplanar, consistent with the assumption that  $h_{ring}/d_{ring}\approx 0.001$ is valid for all of the disk and not just along the chord sampled by the 2007 transit.
The `Thick' model distributes the dust in a flared disk around the substellar object, with a scale height $H$ that increases with distance from the central source $R$ as $H=0.02 R^{1.1}\rm{~au}$.
The temperature structure is computed assuming the disk is optically thin, which provides the brightest possible disk at millimetre wavelengths.
The dust distribution in the  `Flared' model is identical to that in  the Thick model, but the dust is optically thick along the midplane of the disk.
Heating from the primary star J1407 and from the companion J1407b are taken into account using the stellar parameters listed in \citet{vanWerkhoven14} and \citet{Kenworthy15}, and for a range of semi-major separations of the ring system from the central star.

\begin{figure}[tbh]
\begin{center}
\includegraphics[width=0.75\columnwidth]{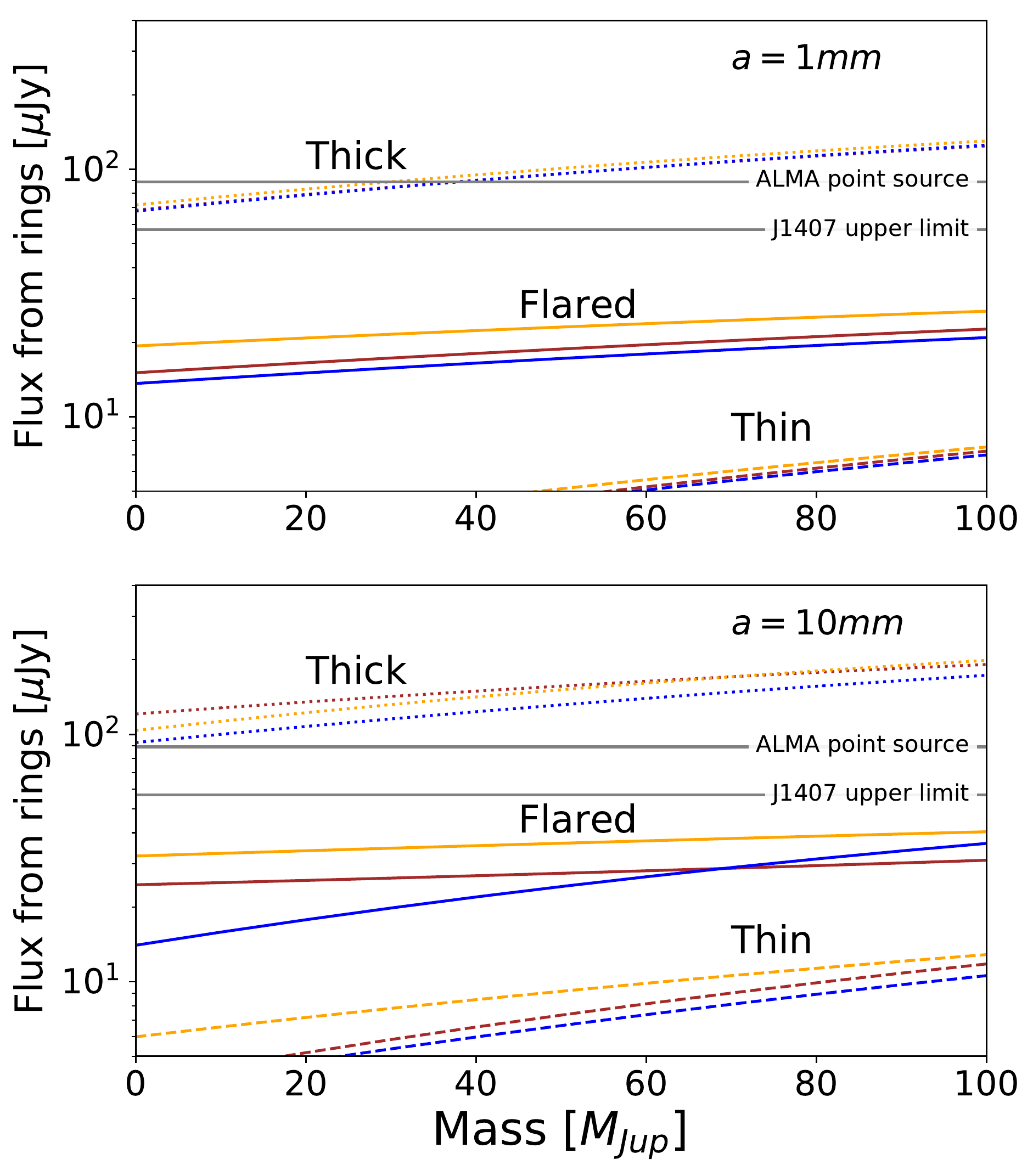}
\caption{\label{fig:fluxmodel}Models for the sub-mm dust emission from the ring system at ALMA Band 7 as a function of mass of the central substellar object using models from \citet{Baraffe08}, together with the ALMA photometry of the point source and the upper limit at the location of the star J1407. Three different disk configurations are presented. Colours correspond to separations from the star J1407: Yellow  5 au, orange  10 au, blue  16 au.}
\end{center}
\end{figure}%

\section{Hypotheses}\label{discuss} 

We discuss four hypotheses to explain our results: 
\begin{itemize}
    \item The ring system is in a bound orbit around the star J1407, and we have an upper limit from the ALMA photometry; 
    \item The detected ALMA source is a ring system that is gravitationally unbound from (and possibly completely unrelated to) the star J1407;
    \item The ALMA source is a background galaxy unrelated to the star or 2007 occulting object; 
    \item The source is a noise peak in the synthesised image. 
\end{itemize}

\subsection{An upper limit on a ring system around J1047} 

If we assume that the ALMA source is not the ring system detected in 2007, then the ring system is in orbit around J1407 and we do not detect it with the ALMA observations.
The models shown in Figure~\ref{fig:fluxmodel} are generated  for different dust geometries and for different separations from the parent star J1407, along with the upper limit at the location of J1407.
The three different colours represent different distances of the ring system from the star:   5, 10, and 15 au for yellow, orange, and blue, respectively.
Neither the Thin nor the Flaring model is constrained by the observations.
The emitted flux increases for larger grain sizes, and we could detect a Thick model disk for all masses above 20 Jupiter masses and particle sizes greater than 1mm.

\subsection{A free-floating ring system} 

We  are looking towards a young stellar association and so there is the possibility that we are seeing a young planet ejected from elsewhere in the association.
The angular separation of the ALMA source from the location of the star $(438\pm8\rm{~mas})$ is consistent with the expected separation for an object that has travelled at 35 \kms{} since the occultation in 2007  ($543\pm82\rm{~mas),}$ which we indicate with the solid white circle in Figure~\ref{fig:alma} and dotted white circles marking the associated uncertainty.
This transverse velocity was derived in \citet{Kenworthy15} by analysis of the steepest gradients in the light curve of the 2007 eclipse and considering a straight-edged opaque occulter moving across the central disk of the star with a velocity perpendicular to the edge of the occulter.
The transverse velocity is then proportional to the rate of change of stellar flux and the diameter of the star;  knowing the light curve gradient and diameter of the star gives the velocity.

The discussion in \citet{Kenworthy15} showed that several gradients would have to be discarded to make a closed circular orbit work for the orbit of the ring system, and instead an eccentric orbit was strongly preferred,  or that the system could be on an unbound hyperbolic orbit with coincidental alignment in 2007.
Looking at the NACO data, any ring system would have a central object with a mass less than $6M_{Jup}$, which would be entirely possible for an unbound ring system since an unbound object does not have an associated Hill sphere.

This could be the first direct detection of the ring system that eclipsed the star J1407 in 2007, and from the models in Figure~\ref{fig:fluxmodel} we can determine that the substellar object has to have a mass $>$6 M$_{Jup}$ for a fixed particle size of 1mm in the Thick model case, or that the particle size is larger than 1mm and the substellar object can be considerably lower in mass.
If this is also the object that occulted J1407 in May 2007, the transverse velocity is 35 \kms{}, which is considerably higher than the escape velocity of the Sco-Cen cluster.
It therefore seems highly unlikely that we are seeing a young planet in the process of being ejected from the cluster.

With this hypothesis, we predict that the ALMA source  has a proper motion of $43$ mas/year, which means that a second ALMA epoch measurement in 2020 will show approximately 130 mas of total proper motion along the line passing through J1407 and the 2017 ALMA source.

\subsection{An unrelated background source} 

Using the galaxy counts measured at 1.3mm with ALMA from \citet{Aravena16} we estimate the source background count at 850 $\mu$m (Band 7) by estimating the number counts at 1.2mm and then converting to 850 $\mu$m number counts.
In \citeauthor{Aravena16}, the flux $S$ in different bands is given as $S_{1.2\rm{mm}}=0.4S_{870\mu m}=0.8S_{1.1\rm{mm}}$.
Taking the mean of the two shorter wavelengths gives $S_{1.2\rm{mm}}=0.6S_{\rm 850\mu m}$.
Our source has $S_{\rm 850\mu m} =0.1{\rm mJy}$ so $S_{1.2\rm{mm}}=0.06 {\rm mJy}$ for our ALMA detection.

Using Table 2 from \citet{Aravena16} we estimate 71500 sources with a flux of $89\rm{~\mu Jy}$ or greater per square degree.
The probability of finding more than zero galaxies within a radius of $0.5$ arcseconds is around 0.0047, or 1 in 212, and the probability of finding a galaxy within the annulus representing a $35\rm{~km/s}$ free-floating object is 0.0029, or 1 in 341.
It seems unlikely that the source is a background galaxy, but we cannot rule this hypothesis out without a second epoch observation with ALMA.

\subsection{A noise fluctuation}

The last hypothesis we consider is that the ALMA source is a noise fluctuation.
This is the brightest peak within the 50\% power point of the synthesised ALMA field of view (and certainly within the proper motion circle).
Our analysis uses the primary beam corrected images, which means the noise increases away from the phase centre, but the detection is close enough to the phase centre that this should not impact the derived statistical significance.
A detection at a second epoch would increase confidence in our interpretation.

\section{Conclusions and future work}\label{concl} 

In this paper we present observations using ALMA and NACO towards the young star J1407 in order to determine the nature of the object that caused the light curve seen in May 2007, and we report the detection of an unresolved source consistent with dust heated by a central substellar object.
We present several hypotheses for the detected ALMA source and we note that it is at an angular separation consistent with a free-floating ring system that happened to move through the line of sight between us and J1407 in May 2007.
The upper limit of $6M_{Jup}$ at the location of the ALMA source implies that we are seeing a free-floating ring system around an exoplanet, assuming the ring system formed at the same time as Sco-Cen.
If the ALMA source is the object which occulted J1407 in 2007, then the proper motion of the ALMA source is $43$ mas/yr. 
An observation with ALMA three or more years after the 2017 detection will show this proper motion with enough significance to confirm or rule out this hypothesis, or it will confirm that the source is a distant background object.
Further searches are being carried out for other large ring systems both in archival data and in current wide field surveys.

\begin{acknowledgements}

  We thank the referee and Editor for comments and suggestions which improved this paper, and several discussions with the ALMA Allegro node in Leiden for understanding the astrometric calibration of ALMA.
This paper makes use of the following ALMA data: ADS/JAO.ALMA\#016.1.00470.S. 
ALMA is a partnership of ESO (representing its member states), NSF (USA) and NINS (Japan), together with NRC (Canada), MOST and ASIAA (Taiwan), and KASI (Republic of Korea), in cooperation with the Republic of Chile.
The Joint ALMA Observatory is operated by ESO, AUI/NRAO and NAOJ.
This research made use of Astropy,\footnote{http://www.astropy.org} a community-developed core Python package for Astronomy \citep{astropy:2013, astropy:2018} and of Matplotlib \citep{Hunter:2007}.
This research made use of APLpy, an open-source plotting package for Python \citep{aplpy2012,aplpy2019}.
This work has made use of data from the European Space Agency (ESA) mission {\it Gaia} (\url{https://www.cosmos.esa.int/gaia}), processed by the {\it Gaia} Data Processing and Analysis Consortium (DPAC, \url{https://www.cosmos.esa.int/web/gaia/dpac/consortium}). Funding for the DPAC has been provided by national institutions, in particular the institutions participating in the {\it Gaia} Multilateral Agreement.
Part of this research was carried out at the Jet Propulsion Laboratory, California Institute of Technology, under a contract with NASA.
\end{acknowledgements}

\bibliographystyle{aa}
\bibliography{Kenworthy_ALMA_J1407}

\end{document}